\documentstyle[11pt,aaspp4,epsf]{article}
\begin{document}
\title{Performance of the optimized Post-Zel'dovich approximation 
for CDM models in arbitrary FLRW cosmologies}
\author{Takashi Hamana}
\affil{Astronomical Institute, Tohoku University, Sendai 980-8578, Japan}
\begin{abstract}
We investigate the performance of the optimized Post-Zel'dovich
approximation in three cold dark matter cosmologies.
We consider two flat models with $\Omega_0=1$ (SCDM) and with
$\Omega_0=0.3$ ($\Lambda$CDM) and an open model with $\Omega_0=0.3$
(OCDM).
We find that the optimization scheme proposed by Wei{\ss}, Gottl\"ober \&
Buchert (1996), in which the performance of the
Lagrangian perturbation theory was optimized only for the Einstein-de Sitter
cosmology, shows the excellent performances not only for SCDM model but
also for both OCDM and $\Lambda$CDM models.
This universality of the excellent performance of the optimized
Post-Zel'dovich approximation is explained by the fact that a relation 
between the Post-Zel'dovich order's growth factor $E(a)$ and Zel'dovich
order's one $D(a)$, $E(a)/D^2(a)$, is insensitive to the background
cosmologies.
\end{abstract}

\section{Introduction}
The Zel'dovich approximation (Zel'dovich, 1970) is known to be
accurate even in the weakly non-linear regime of the structure
formation. 
However, when the shell-crossing occurs, the validity of the
approximation breaks down.
Coles, Melott \& Shandrin (1993) introduced a truncated Zel'dovich
approximation by smoothing the small-scale power in the initial
conditions to remove the unwanted non-linearity.
Optimization of the smoothing schemes, the filter shapes and filter
scales, and the performance of the approximation have been
investigated by authors (Melott, Pellman \&
Shandrin, 1994, Melott, Buchert \& Wei\ss, 1995).
They found that the truncated Zel'dovich
approximation does not provide a correct description of the internal
structure and mass distribution of non-linear structures like galaxy
clusters, but it is accurate in locating their positions and
thus reliably describes their spatial distribution.
Therefore, the truncated Zel'dovich approximation is a powerful tool to 
study the large-scale distribution of galaxy clusters 
which provides important constraints on models of cosmic
structure formation (see e.g. Borgani et al., 1995).

From a theoretical point of view, very recently, Takada \& Futamase
(1998) developed a formalism which allows one to investigate a
relation between the large-scale quasi non-linear dynamics and the 
small-scale non-linear dynamics within using an averaging method in 
the Lagrangian perturbation theory.
They found that the small-scale dynamics only weakly
affects the large-scale structure formation and thus the truncated
Lagrangian perturbation theory is a good approximation to investigate 
the large-scale structure formation.

The second-order correction to the truncated Zel'dovich approximation
was introduced by Melott (1994), and Melott et al.\ (1995). 
They found that the second-order correction improves the performance
of the approximation.
In their study, Wei{\ss} et al.\ (1996) optimized the performance of
the truncated second-order Zel'dovich approximation in the
Einstein-de Sitter cosmology.
They performed N-body simulations of the cold dark matter (CDM)
and broken scale invariance models, and compared the results with
those obtained from the optimized approximation scheme.
They found an excellent performance of the optimized approximation
down to scales close to the correlation length.
However, since the optimization was performed only in the Einstein-de
Sitter cosmology, 
it was not clear whether the same optimized scheme would perform as
excellent in arbitrary Friedmann-Lema\^{\i}tre-Robertson-Walker
(FLRW) cosmologies.
Since the optimized approximation is a powerful tool to investigate
the formation of large-scale structures,
it is worth while generalizing the optimized scheme of the
approximation to arbitrary FLRW cosmologies. 

The main purpose of this paper is to test the performance of the
optimized second-order Zel'dovich approximation obtained by Wei{\ss}
et al.\ (1996) (hereafter, optimized Post-Zel'dovich approximation) in 
arbitrary FLRW cosmologies.
We examine three CDM models, two flat models with
$\Omega_0=1$ (SCDM) and with $\Omega_0=0.3$ ($\Lambda$CDM) and an open
model with $\Omega_0=0.3$ (OCDM). From the results of the optimized 
Post-Zel'dovich approximation, we calculate the two-point correlation 
functions.
We do not perform the N-body simulation, but we compare the
correlation functions with those predicted by a parameterized fitting
formula that Peacock \& Dodds (1996) use to predict the power spectrum 
of the non-linear mass density field.
We refer the reader to the above reference for details of the formula
and its implementation.
In their recent paper, Jenkins et al.\ (1998) compared the two-point
correlation functions obtained from their very large N-body simulations
with that predicted by fitting formula by Peacock \& Dodds (1996) and 
found a good agreement between them over a scales between 
$\sim 0.1h^{-1}$Mpc and $\sim 10 h^{-1}$Mpc.

The plan of this paper is as follows.
Section 2 gives a brief summary of the optimized Post-Zel'dovich
approximation. 
The performance of the approximation in arbitrary FLRW cosmologies is
tested in section 3. 
Our paper concludes in section 4 with discussions.

Throughout this paper, we use a unit for which $c=1$, and the scale
factor $a$ is normalized to unity at the present epoch, i.e.,
$a_0=1$.
The Hubble parameter $H$, density parameter $\Omega$ and normalized
cosmological constant $\lambda$ are defined in the usual manner.
Quantities of the present epoch and initial epoch are indicated by
indices $0$ and $i$, respectively. 

\section{Summary of the optimized Post-Zel'dovich approximation}
The Zel'dovich and Post-Zel'dovich approximation are regarded as
subclasses of the first order and second
order solution of the Lagrangian perturbation theory, respectively
(Buchert, 1989, 1992).
Since the Lagrangian perturbation theory has been 
thoroughly investigated by various authors (e.g., Buchert \& Ehlers, 1993,
Buchert, 1994, Bouchet et al., 1995, Sasaki \& Kasai, 1998), 
here we describe only the
aspects which are directly relevant to this paper.

Denoting the comoving Eulerian coordinates by ${\bf{x}}$, and Lagrangian 
coordinates 
by ${\bf{q}}$, the field of trajectories ${\bf{x}} = {\bf{F}}
({\bf{q}},a)$ up to the Post-Zel'dovich order is
\begin{equation}
\label{PZAform}
{\bf{x}} = {\bf{q}} + D(a) \nabla \Psi^{(1)} + E(a) \nabla \Psi^{(2)},
\end{equation} 
with the time-dependent coefficients expressed in terms of the linear
growth rate $D_+(a)$ (Peebles, 1980)  
\begin{equation}
\label{ZAtime}
D(a)={{D_+(a)} \over {D_+(a_i)}} - 1, 
\end{equation}
\begin{equation}
\label{PZAtime}
E(a) = {{3 \Omega_0 {H_0}^2} \over 4} H(a) 
\int^a {{da'} \over {(H(a')a')^3}}
\int^{a'} {da''} {{D^2 (a'')} \over {{a''}^2}},
\end{equation}
where $H(a)$ is the Hubble parameter,
$H(a) \equiv H_0 ( \Omega_0/{a^3} +\lambda_0 - K/a^2)^{1/2}$.
In the above expressions, we only take the fastest growing mode for
each order.
It is important to note that the relation between $D(a)$ and $E(a)$,
$E(a)/D^2(a)$, is remarkably insensitive to the background cosmologies 
(Bouchet et al., 1992, 1996).
The displacement potentials are obtained by solving iteratively
two Poisson equations;
\begin{equation}
\label{pois1}
\Delta \Psi^{(1)} = -\delta_i,
\end{equation}
\begin{equation}
\label{PZAspa}
\Delta \Psi^{(2)}= \Psi,_{jk}^{(1)} \Psi,_{kj}^{(1)} 
- \Psi,_{jj}^{(1)} \Psi,_{kk}^{(1)},
\end{equation}
where $\delta_i$ is an initial density contrast field.

Next, we review the optimization scheme proposed by
Wei{\ss} et al.\ (1996), which we adopt in this paper.
The high frequency part of the Fourier transform of an initial density 
field is smoothed out by a Gaussian $k$-space filter with a
characteristic smoothing scale $k_{gs}$,
\begin{equation}
\label{wgs}
W(k) = \exp \left( - {{k^2} \over {2 k_{gs}^2}} \right),
\end{equation}
i.e., $P(k) \rightarrow P(k) W^2(k)$, where $P(k)$ is the power
spectrum of the initial density field.
Wei{\ss} et al.\ (1996) found that an optimal value of $k_{gs}$ does
not significantly depend on the form of $P(k)$, and is related to a
scale of non-lineality $k_{nl}$ by $k_{gs}  \sim 1.2 k_{nl}$
with a little scatter. 
The quantity $k_{nl}$ is defined by
\begin{equation}
\label{knl}
{{D_+^2(a)} \over {(2\pi)^3}} \int_0^{k_{nl}} d^3 k P(k) =1.
\end{equation} 
We adopt the recommended value of $k_{gs} = 1.2 k_{nl}$.

\section{Models and results}
We examine three CDM models.
Table 1 lists the models and gives their parameters.
We use the CDM transfer function in Bardeen et
al.\ (1986), with the scale invariant ($n=1$) primordial power spectrum.
The shape parameter, $\Gamma$, in the spectrum defined by Sugiyama
(1995), which we adopt, is 
\begin{equation}
\label{Gamma}
\Gamma = \Omega_0 h \exp \left[ -\Omega_b \left(1+ \sqrt{2 h}/
\Omega_0 \right) \right],
\end{equation}
where $\Omega_b$ is the baryonic matter density parameter, and $h$ is
the normalized Hubble constant, i.e., $H_0 = 100 h$km/sec/Mpc.
In all cases, the amplitude of primordial fluctuation is set such that
the models reproduce the observed abundance of rich galaxy clusters of 
the present day. 
We adopt the values of $\sigma_8$ recommended by Eke, Cole \& Frenk,
(1996).

\begin{table}[t]
\caption{Summary of model parameters.\label{table1}}
\begin{tabular}{cccccc}
Model & $\Omega_0$  & $\lambda_0$ & $\Omega_b h^2$ & $h$ & $\sigma_8$ \\
SCDM & 1.0 & 0.0 & 0.015 & 0.5 & 0.51 \\ 
OCDM & 0.3 & 0.0 & 0.015 & 0.7 & 0.85 \\ 
$\Lambda$CDM & 0.3 & 0.7 & 0.015 & 0.7 & 0.90 \\ 
\end{tabular}
\end{table}

The initial density field is set on $128^3$ grid points for a cubic
box of $L=128 \times \Delta x h^{-1}$Mpc a side, with a periodic
boundary condition, where $\Delta x$ is the grid spacing.
Here we use Bertschinger's software COSMICS (Bertschinger, 1995) with some 
modifications according to the shape parameter (\ref{Gamma}), and
smoothing of the power spectrum of the initial density field with
the Gaussian filter (\ref{wgs}).
In order to ensure the condition $\delta_i \ll 1$, which is required
for deriving eq.\ (\ref{pois1}), the initial
condition is set at a redshift $z_i = 10^3$.
We solve the Poisson equations (\ref{pois1}) and (\ref{PZAspa}) via Fast
Fourier Transformation.
Then we move $128^3$ particles having the initial Lagrangian coordinate on 
the grid, according to the Post-Zel'dovich approximation,
eq.\ (\ref{PZAform}).
Here, for the linear growth factor $D_+ (a) $, we use the fitting
formula of Carrol, Press \& Turner (1992), and the Post-Zel'dovich
order's growth rate $E(a)$ is evaluated by numerical integrations. 

We consider three cases of the boxsize. The grid spacings and 
particle masses are summarized in Table 2.
The redshifts of realizations are chosen to be $z=0$, $1$ and $2$, 
and the scales of non-linearity, $k_{nl}$, for each redshift for each
model are presented in Table 3. 

\begin{table}[t]
\caption{The grid spacing $\Delta x$ and particle mass.\label{table2}}
\begin{tabular}{cccc}
Model & Label  & $\Delta x$ ($h^{-1}$Mpc) & Particle mass
($h^{-1}M_{\odot}$)\\ 
SCDM & S & 0.5 & $3.47\times 10^{10}$ \\
{} & M & 1.0 & $2.78\times 10^{11}$ \\
{} & L & 2.0 & $2.22\times 10^{12}$ \\
OCDM & S & 0.75 & $3.51\times 10^{10}$ \\
and & M & 1.5 & $2.81\times 10^{11}$ \\
$\Lambda$CDM & L & 3.0 & $2.25\times 10^{12}$\\ 
\end{tabular}
\end{table}
%
\begin{table}[t]
\caption{The scale of non-linearity $k_{nl}$
($h$Mpc${}^{-1}$).\label{table3}}
\begin{tabular}{cccc}
Redshift of realization & SCDM  & OCDM & $\Lambda$CDM\\ 
$z=0$ & 0.548  & 0.303 & 0.279 \\
$1$ & 1.59 & 0.559 & 0.605 \\
$2$ & 3.62 & 0.933 & 1.26 \\
\end{tabular}
\end{table}

From the results of the realizations, we evaluate the two-point
correlation function of the particles by adopting the direct estimator 
(Hockney \& Eastwood, 1988),
\begin{equation}
\label{2cor}
\xi (r) = {{N_p \Delta x^3} \over {N_c \delta V}} -1,
\end{equation}
where $N_p$ is the number of pairs of particles with separations
between $r-\Delta /2$ and $r+\Delta /2$, $\delta V$ is the volume of
this shell and $N_c$ is the number of particles taken as centers. 
The results are plotted in Figure 1.

\begin{figure}
\begin{center}
\begin{minipage}{7.5cm}
\begin{center} 
\epsfxsize=7.5cm
\epsffile{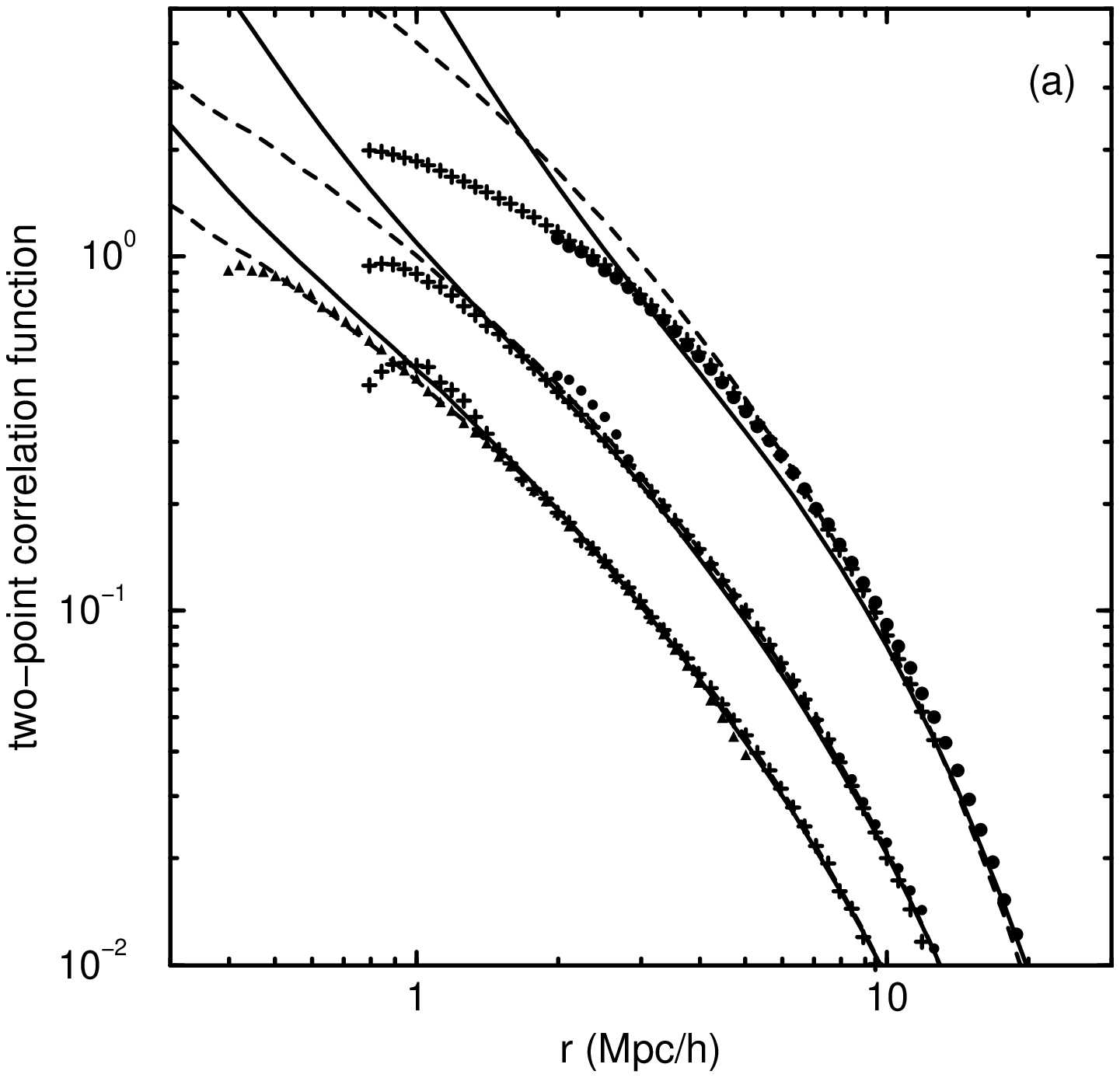}
\end{center}
\end{minipage}
\hspace{0.5cm}
\begin{minipage}{7.5cm}
\begin{center} 
\epsfxsize=7.5cm
\epsffile{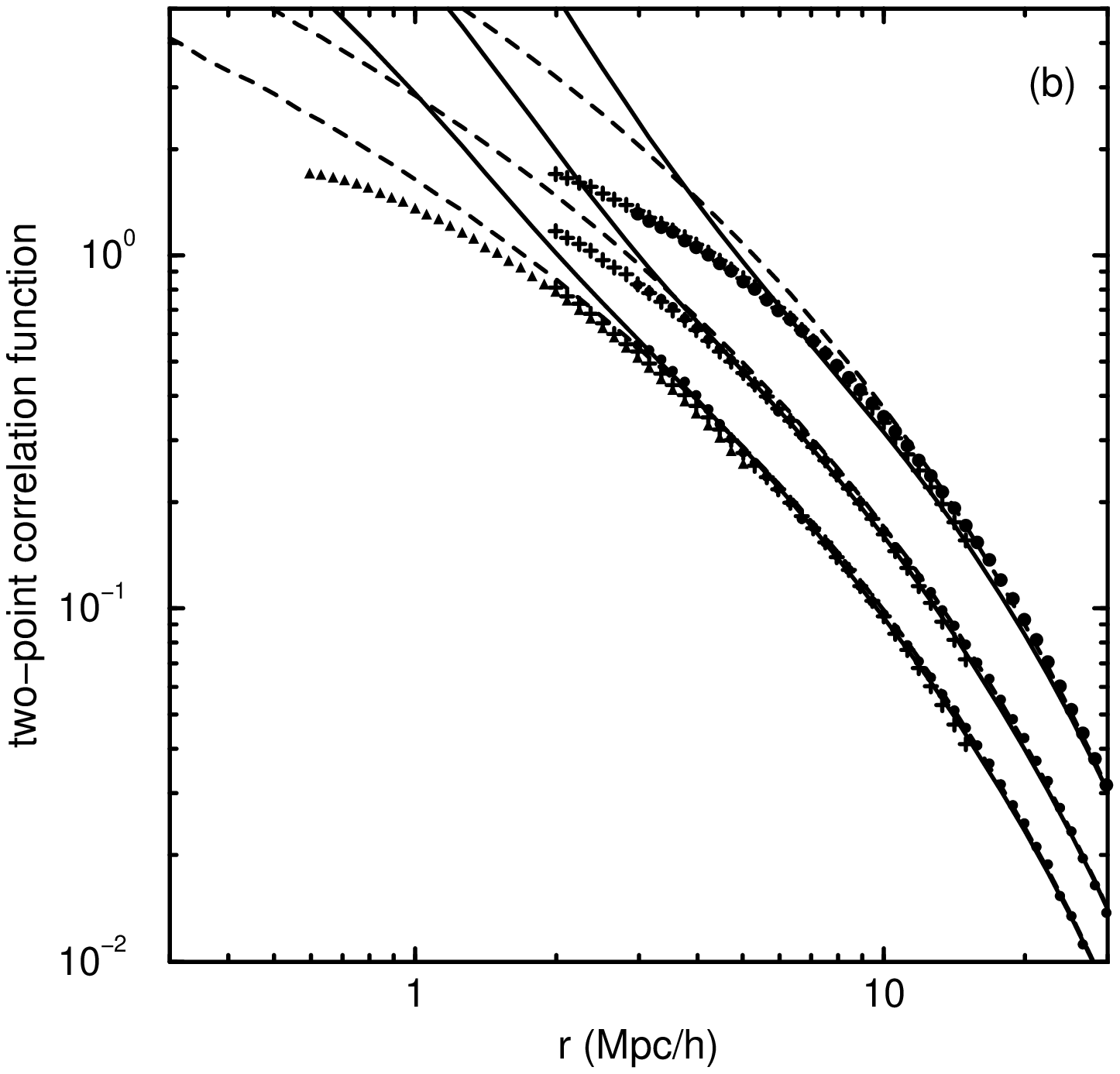}
\end{center}
\end{minipage}
\begin{minipage}{7.5cm}
\begin{center} 
\epsfxsize=7.5cm
\epsffile{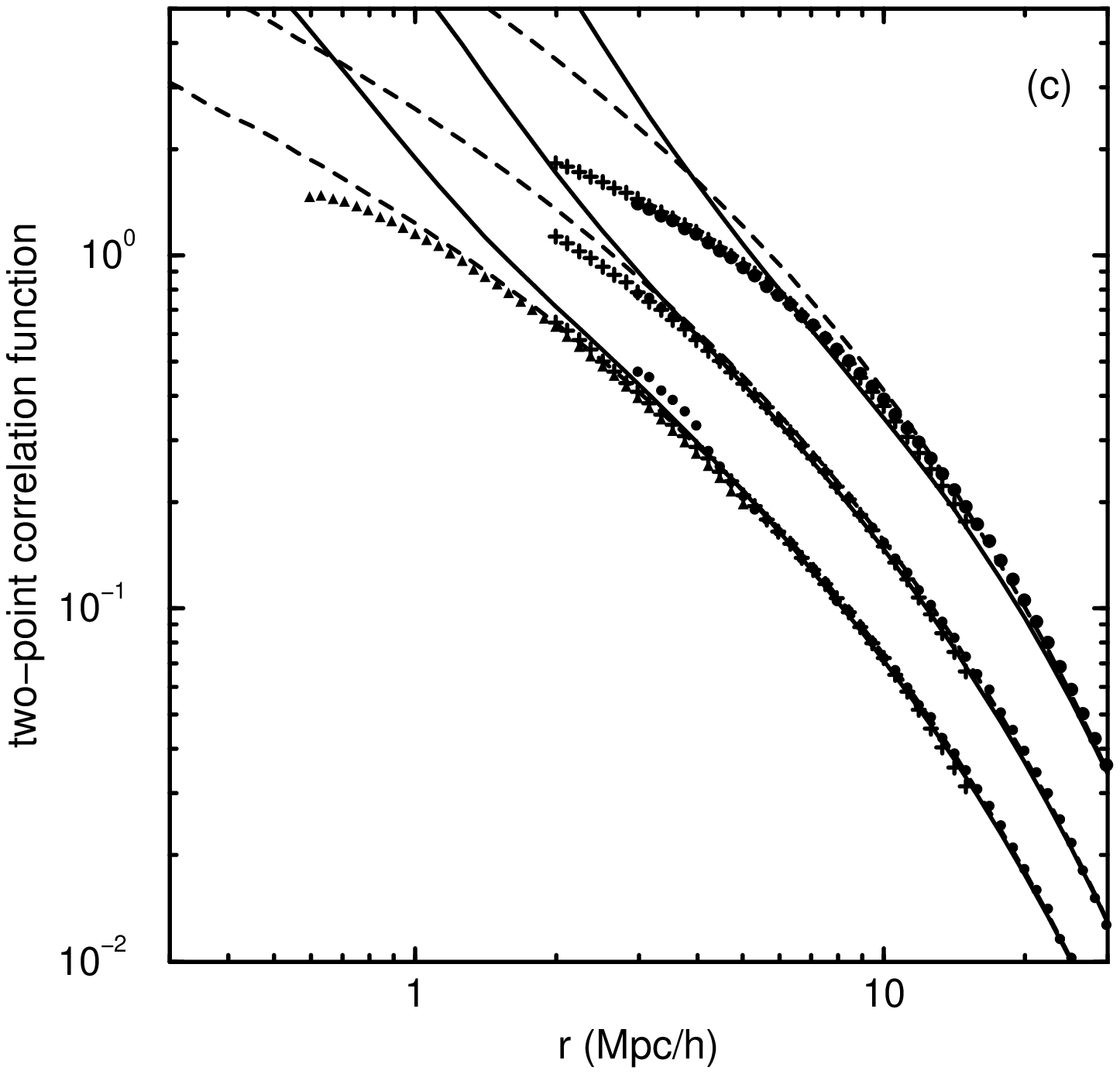}
\end{center}
\end{minipage}
\hspace{0.5cm}
\begin{minipage}{7.5cm}
\begin{center} 
\caption[hamana1a.ps,hamana1b.ps,hamana1c.ps]{
Two-point correlation functions compared with those
predicted by the linear theory and fitting formula by Peacock \& Dodds 
(1996).
The results of our S, M and L box realizations are shown by filled
triangles, pluses and filled circles, respectively.
Solid lines show the correlation function derived from the
non-linear power spectrum, and dashed lines  are those predicted by
the linear theory.
(a) is for SCDM model, (b) is for OCDM model and (c) is for
$\Lambda$CDM model.
In all panels, from the top to bottom, the redshifts of
realizations are $z=0$, $1$ and $2$, respectively. \label{fig1}}
\end{center}
\end{minipage}
\end{center}
\end{figure}

It can be shown in Figure 1a that our results for SCDM model, of course,
agree well with that obtained by Wei{\ss} et al.\ (1996) (see Figure 7 of
their paper). 
It can be also shown in Figure 1b and 1c that the comparable
performance of the optimized Post-Zel'dovich approximation is also
achieved in both OCDM and $\Lambda$CDM models.
In all cases, the correlation functions obtained from the
approximation are accurate down to the
scales where the non-linear correlation functions represent their
non-linear behavior, and below those scales they are depressed.
Meanwhile the correlation length itself is underestimated only slightly by 
less than $1h^{-1}$Mpc (this point has been also pointed out by
Wei{\ss} et al.\ 1996).
For the high redshift realization cases, the correlation functions 
agree well with those predicted by the linear theory down to very
close to the correlation length.

\section{Discussions}
In the last section, we found that the optimized Post-Zel'dovivh
approximation by Wei{\ss} et al.\ (1996) performs excellently not only
in SCDM model but also in both OCDM and $\Lambda$CDM models.
This is, apriori, not clear because in Wei{\ss} et al.\ (1996) the
optimization was performed only in the Einstein-de Sitter background
cosmology. 

The universality of the excellent performance of the optimized
Post-Zel'dovich approximation may be explained as follows.
The smoothing scale $k_{gs}$ is chosen to be related to the scale of
non-linearity $k_{nl}$, and $k_{nl}$ is determined by the integral of
the linearly evolved power spectrum of the initial density field,
eq.\ (\ref{knl}), i.e., the power spectrum evolves as $\propto D_+^2(a) 
\simeq D^2(a)$.
On the other hand, the Post-Zel'dovich order's displacement is
evolved according to $E(a)$ and the unwanted non-linearity in the
initial data is
removed by the Gaussian filter with the smoothing scale $k_{gs}$.
Thus, $k_{nl}$ relates to $D^2(a)$, and $k_{gs}$ relates to $E(a)$.
Therefore, an optimized relation between $k_{gs}$ and $k_{nl}$ is
determined by the relation between $D^2(a)$ and $E(a)$.
As was pointed out by Bouchet et al.\ (1992) and (1996), $E(a)/D^2(a)$
is very insensitive to the cosmologies.
Therefore, once an optimized relation between $k_{gs}$ and $k_{nl}$
is found in a background cosmology, it is also optimized in arbitrary
background cosmologies.
We should also note that although we dealt only with the CDM model, 
it has been found that the optimized relation
between $k_{gs}$ and $k_{nl}$ is insensitive to shapes of the power
spectrum (Melott et al., 1994, 1995, Wei{\ss} et al., 1996). 

From a computational point of view, the optimized Post-Zel'dovich
approximation is very time-efficient;
this enables us to compute many
independent models within a reasonable CPU time.
Although, the approximation can not provide a correct description of
the internal structure of non-linear structures, it describes their
spatial distribution well (Wei{\ss} et al., 1996).
As was shown in the last section, two-point correlation functions are
accurate down to the scale close to the correlation length.
Therefore the approximation is a powerful tool to study the formation
of large-scale structure on scales above the correlation length, such
as the analysis of the cluster distribution (Borgani et al., 1995).
It is also appropriate for the study of the gravitational lensing by
the large-scale structures (Bertelmann \& Schneider, 1991).

Before closing this paper, we propose to use the Post-Zel'dovich
approximation for setting an initial condition of N-body simulations.
The Post-Zel'dovich order's solutions can be easily obtained from the
Zel'dovich solutions, eqs.\ (\ref{PZAtime}) and (\ref{PZAspa}).
Thus, the Post-Zel'dovich approximation is as easy to implement as the
Zel'dovich approximation.
Therefore, one who adopts the Zel'dovich approximation to set the
initial condition can easily include the
Post-Zel'dovich correction, which will improve the accuracy of the 
simulations.

\acknowledgments

The author would like to thank M.\ Kasai for valuable
comments on this paper, M.\ Morita for useful discussion on the
Zel'dovich approximation and P.\ Premadi for carefully reading and
commenting the manuscript. 
He also would like to thank T.\ Futamase and M.\ Takada for providing
their manuscript prior to publication and for useful discussion.


%
\end{document}